\documentstyle[12pt]{article}                    %%%
\newfont{\ffont}{msym10}                          %%
\newcommand{\beq}{\begin{equation}}               %%
\newcommand{\eeq}{\end{equation}}                 %%
\newcommand{\bqry}{\begin{eqnarray}}              %%
\newcommand{\eqry}{\end{eqnarray}}                %%
\newcommand{\bqryn}{\begin{eqnarray*}}            %%
\newcommand{\eqryn}{\end{eqnarray*}}              %%
\newcommand{\NL}{\nonumber \\}                    %%
\newcommand{\preprint}[1]{\begin{table}[t]        %%
            \begin{flushright}                    %%
            \begin{large}{#1}\end{large}          %%
            \end{flushright}                      %%
            \end{table}}                          %%
\newcommand{\PD}[2]                               %%
    {\frac{\partial^{#2}}{\partial #1^{#2}}}      %%
\begin{document}
\preprint{LA-UR-97-717}
\title{Towards Resolution of the Enigmas of \\ $P$-Wave Meson  
Spectroscopy}
\author{\\ L. Burakovsky\thanks{E-mail: BURAKOV@PION.LANL.GOV} \
and \ T. Goldman\thanks{E-mail: GOLDMAN@T5.LANL.GOV} \
\\  \\  Theoretical Division, MS B285 \\  Los Alamos National  
Laboratory \\
Los Alamos, NM 87545, USA \\}
\date{ }
\maketitle
\begin{abstract}
The mass spectrum of $P$-wave mesons is considered in a nonrelativistic 
constituent quark model. The results show the common mass degeneracy of 
the isovector and isodoublet states of the scalar and tensor meson  
nonets, and
do not exclude the possibility of a similar degeneracy of the same  
states of
the axial-vector and pseudovector nonets. Current experimental  
hadronic and
$\tau$-decay data suggest, however, a different scenario leading to  
the $a_1$
meson mass $\simeq 1190$ MeV and the $K_{1A}$-$K_{1B}$ mixing angle  
$\simeq (
37\pm 3)^o.$ Possible $s\bar{s}$ states of the four nonets are also  
discussed.
\end{abstract}

\bigskip
\section{Introduction}
The existence of a gluon self-coupling in QCD suggests that, in  
addition to
the conventional $q\bar{q}$ states, there may be non-$q\bar{q}$  
mesons: bound
states including gluons (gluonia and glueballs, and $q\bar{q}g$  
hybrids) and
multiquark states \cite{1}. Since the theoretical guidance on the  
properties
of unusual states is often contradictory, models that agree in the  
$q\bar{q}$
sector differ in their predictions about new states. Among the naively 
expected signatures for gluonium are \hfil\break
i) no place in $q\bar{q}$ nonet, \hfil\break
ii) flavor-singlet coupling, \hfil\break
iii) enhanced production in gluon-rich channels such as $J/\Psi  
(1S)$ decay,
\hfil\break iv) reduced $\gamma \gamma $ coupling, \hfil\break v) exotic 
quantum numbers not allowed for $q\bar{q}$ (in some cases). \hfil\break
Points iii) and iv) can be summarized by the Chanowitz $S$  
parameter \cite{Cha}
$$S=\frac{\Gamma (J/\Psi (1S)\rightarrow \gamma X)}{{\rm PS} (J/\Psi (1S)
\rightarrow \gamma X)}\times \frac{{\rm PS} (X\rightarrow \gamma  
\gamma )}{
\Gamma (X\rightarrow \gamma \gamma )},$$
where PS stands for phase space. $S$ is expected to be larger for  
gluonium
than for $q\bar{q}$ states. Of course, mixing effects and other  
dynamical
effects such as form-factors can obscure these simple signatures.  
Even if the
mixing is large, however, simply counting the number of
observed states remains a clear signal for non-exotic  
non-$q\bar{q}$ states.
Exotic quantum number states $(0^{--},0^{+-},1^{-+},2^{+-},\ldots  
)$ would be
the best signatures for non-$q\bar{q}$ states. It should be also  
emphasized
that no state has yet unambiguously been identified as gluonium, or as a 
multiquark state, or as a hybrid.

In this paper we shall discuss $P$-wave
meson states, the interpretation of which as members of  
conventional quark
model $q\bar{q}$ nonets encounters difficulties \cite{enigmas}. We  
shall be
concerned with the scalar, axial-vector,
pseudovector and tensor meson nonets which have the following  
$q\bar{q}$ quark
model assignments, according to the most recent Review of
Particle Physics \cite{pdg}:\hfil\break
1) $\;1\; ^1P_1$ pseudovector meson nonet, $J^{PC}=1^{+-},$
$b_1(1235),\!\;h_1(1170),\!\;h_1^{'}(1380),\!\;K_{1B}$ $\! ^1$\hfil\break
2) $\;1\; ^3P_0$ scalar meson nonet, $\;\;J^{PC}=0^{++},\;\;\;$  
$a_0($ ? $),
\;\;\;f_0($ ? $),\;\;\;f_0^{'}($ ? $),\;\;\;K_0^\ast (1430)$\hfil\break
3) $\;1\; ^3P_1$ axial-vector meson nonet, $J^{PC}=1^{++},\;$
$a_1(1260),\;f_1(1285),\;f_1^{'}(1510),\;K_{1A}$ \footnote{The  
$K_{1A}$ and
$K_{1B}$ are nearly $45^o$ mixed states of the $K_1(1270)$ and  
$K_1(1400)$
\cite{pdg}, their masses is therefore $\simeq 1340$ MeV.}\hfil\break
4) $\;1\; ^3P_2$ tensor meson nonet, $\;J^{PC}=2^{++},\;$
$a_2(1320),\;f_2(1270),\;f_2^{'}(1525),\;K_2^\ast (1430),$ and  
start with a
review of the states and their possible masses and assignments, as  
viewed by
several groups.
 \\   \\
1. Scalar meson nonet.\hfil\break
The spectrum of the scalar meson nonet is a long-standing problem  
of light
meson spectroscopy. The number of resonances found in
the region of 1--2 GeV exceeds the number of states that  
conventional quark
models can accommodate \cite{pdg}. Extra states are interpreted  
alternatively
as $K\bar{K}$ molecules, glueballs, multi-quark states or hybrids. In 
particular, except for a well established scalar isodoublet state, the 
$K_0^\ast (1430),$ the Particle Data Group (PDG) \cite{pdg} lists two 
isovector states, the $a_0(980)$ and $a_0(1450).$ The latter,  
having mass and
width 1450$\pm 40$ MeV, $270\pm 40$ MeV, respectively, was  
discovered recently
by the Crystal Barrel collaboration at LEAR \cite{LEAR}. The third  
isovector
state (not included in \cite{pdg}), $a_0(1320),$ having mass and  
width $1322
\pm 30$ MeV and $130\pm 30$ MeV, was seen by GAMS \cite{GAMS} and LASS 
\cite{Aston} in the partial wave analyses of the $\eta \pi $ and  
$K_sK_s$ data,
respectively. There are four isoscalar states in \cite{pdg}, the
$f_0(400-1200)$ (or $\sigma ),$ the interpretation of which as a  
particle is
controversial due to a huge width of 600--1000 MeV, $f_0(980),$  
$f_0(1370)$
(which stands for two separate states, $f_0(1300)$ and $f_0(1370)$, of a 
previous edition of PDG \cite{pdg1}), and $f_0(1500)$ (which also  
stands for
two separate states, $f_0(1525)$ and $f_0(1590),$ of a previous  
edition of
PDG), and two more possibly scalar states, the $f_J(1710),$ $J=0$  
or 2, seen
in radiative $J/\Psi $ decays, and an $\eta $-$\eta $ resonance  
$X(1740)$ with
uncertain spin, produced in $p\bar{p}$ annihilation in flight
and in charge-exchange. Recently several groups claimed different scalar 
isoscalar structures close to 1500 MeV, including a narrow state  
with mass
$1450\pm 5$ MeV and width $54\pm 7$ MeV \cite{Abatzis}. The  
lightest of the
three states at 1505 MeV, 1750 MeV and 2104 MeV revealed upon  
reanalyzing of
data on $J/\Psi \rightarrow \gamma 2\pi ^{+}2\pi ^{-}$ \cite{Bugg},  
and the
$f_0(1450),$ $f_0(1500),$ $f_0(1520).$ The masses, widths and decay  
branching
ratios of these states are incompatible within the errors quoted by  
the groups.
We do not consider it as plausible that so many scalar isoscalar  
states exist
in such a narrow mass interval. Instead, we take the various states as 
manifestation of one object which we identify with the $f_0(1525)$ of a
previous edition of PDG.

It has been convincingly argued that the narrow $a_0(980)$, which  
has also
been seen as a narrow structure in $\eta \pi $ scattering, can be  
generated by
meson-meson dynamics alone \cite{WI,KK}. This interpretation of the  
$a_0(980)$
leaves the $a_0(1320)$ or $a_0(1450)$ (which may be manifestations  
of one
state having a mass in the interval 1350-1400 MeV) as the 1 $^3P_0$  
$q\bar{q}$
state. Similarly, it is mostly assumed that the $f_0(980)$ is a  
$K\bar{K}$
molecule, as suggested originally by Weinstein and Isgur \cite{WI}.
The mass degeneracy and their proximity to the $K\bar{K}$ threshold 
seem to require that the nature of both, $a_0(980)$ and $f_0(980),$  
states
should be the same. On the other hand, the $K\bar{K}$ interaction  
in the $I=1$
and $I=0$ channels is very different: the extremely attractive $I=0$ 
interaction may not support a loosely bound state. Instead, it may  
just define
the pole position of the $f_0(980)$ $q\bar{q}$ resonance. Indeed, a  
recent
analysis of all high statistics data in the neighborhood of the  
$K\bar{K}$
threshold done by Pennington \cite{enigmas}, indicates in an almost 
model-independent way that the $f_0(980)$ is not a $K\bar{K}$ molecule. 
Moreover, Morgan and Pennington \cite{MP} find the $f_0(980)$ pole  
structure
characteristic for a genuine resonance of the constituents and not  
of a weakly
bound system. The $I=1$ $K\bar{K}$ interaction is weak and may  
generate a
$K\bar{K}$ molecule. Alternatively, T\"{o}rnqvist \cite{Torn}
interprets both the $f_0(980)$ and $a_0(980)$ as the members of the  
$q\bar{q}$
nonet with strong coupling to the decay channels, which, however,  
does not
account for the recently discovered $a_0(1320)$ and $a_0(1450).$

With respect to the $f_0(1370)$ (or two separate states,  
$f_0(1300)$ and $f_0(
1370),$ according to a previous edition of PDG), we follow the  
arguments of
Morgan and Pennington \cite{MP} and assume that the $\pi \pi $  
interaction
produces both very broad, $f_0(1000),$ and narrow, $f_0(980),$  
states, giving
rise to a dip at 980 MeV in the squared $\pi \pi $ scattering  
amplitude $T_{11
}.$ In this picture, the $f_0(1370)$ is interpreted as the  
high-mass part of
the $f_0(1000)$ (the low-mass part may be associated with the  
$\sigma $ of
recent PDG). In experiments, the $f_0(1000)$ shows up at $\sim 1300$ MeV 
because of the pronounced dip in $|T_{11}|^2$ at $\sim 1$ GeV. The  
$f_0(1000)$
has an extremely large width; thus the resonance interpretation is
questionable. It could be generated by $t$-channel exchanges instead of 
inter-quark forces \cite{ZB}.

The $f_0(1500)$ state has a peculiar decay pattern \cite{AC}
\beq
\pi \pi :\eta \eta :\eta \eta ^{'}:K\bar{K}=1.45:0.39\pm  
0.15:0.28\pm 0.12:
<0.15.
\eeq
This pattern can be reproduced by assuming the existence of a  
further scalar
state which is mainly $s\bar{s}$ and should have a mass of about  
1700 MeV,
possibly the $f_J(1710),$ and tuning the mixing of the $f_0(1500)$  
with the
$f_0(1370)$ $n\bar{n}$ and the (predicted) $f_0(1700)$ $s\bar{s}$ states 
\cite{AC}. In this picture, the $f_0(1500)$ is interpreted as a  
glueball state
with strong mixing with the close-by conventional scalar mesons. An 
interpretation of the $f_0(1500)$ as a conventional $q\bar{q}$  
state, as well
as a qualitative explanation of its reduced $K\bar{K}$ partial  
width, were
given by Klempt {\it et al.} \cite{Klempt} in a relativistic quark  
model with
linear confinement and an instanton-induced interaction. A quantitative 
explanation of the reduced $K\bar{K}$ partial width of the  
$f_0(1500)$ was
given in a very recent publication by the same authors \cite{Klempt1}.

The above arguments lead one to the following spectrum of the  
scalar meson
nonet (in the order: isovector, isodoublet, isoscalar mostly singlet,
isoscalar mostly octet),
\beq
a_0(1320)\;\;{\rm or}\;\;a_0(1450),\;\;\;K_0^\ast  
(1430),\;\;\;f_0(980)\;\;
{\rm or}\;\;f_0(1000),\;\;\;f_0(1500).
\eeq
This spectrum agrees essentially with the $q\bar{q}$ assignments  
found by
Klempt {\it et al.} \cite{Klempt}, and Dmitrasinovic \cite{Dmitra} who
considered the Nambu--Jona-Lasinio model with a $U_A(1)$ breaking
instanton-induced 't Hooft interaction. The spectrum of the
meson nonet given in \cite{Klempt} is
\beq
a_0(1320),\;K_0^\ast (1430),\;f_0(1470),\;f_0(980),
\eeq
while that suggested by Dmitrasinovic, on the basis of the sum rule
\beq
m_{f_0}^2+m_{f_0^{'}}^2+m_\eta ^2+m_{\eta  
^{'}}^2=2(m_K^2+m_{K_0^\ast }^2)
\eeq
derived in his paper, is \cite{Dmitra}
\beq
a_0(1320),\;K_0^\ast (1430),\;f_0(1590),\;f_0(1000).
\eeq
The $q\bar{q}$ assignment obtained by one of the authors by the  
application of
the linear mass spectrum discussed in ref. \cite{linear} to a  
composite system
of the two, pseudoscalar and scalar nonets, is \cite{invited}
\beq
a_0(1320),\;K_0^\ast (1430),\;f_0(1525),\;f_0(980),
\eeq
in essential agreement with (3) and (5).
 \\   \\
2. Axial-vector meson nonet.\hfil\break
1) One of the uncertainties related to the axial-vector nonet is  
the still
undefined properties of its $I=1$ member, the $a_1(1260)$ meson.  
This meson
has a huge width of $\sim 400$ MeV, due to strong coupling to a  
dominant decay
channel $a_1\rightarrow \rho \pi ,$ which makes the determination  
of its mass
rather difficult. A decade ago Bowler \cite{Bow} argued that,  
according to
his parametrization of the couplings as functions of mass, the  
$\tau $-decay
and hadronic data as of 1986 were entirely consistent as far as the $a_1$
mass was concerned, and concluded that the $a_1$ mass and width are  
safely
within the ranges $\simeq 1235\pm 40$ MeV and $400\pm 100$ MeV,  
respectively.
These values are in excellent agreement with those currently  
adopted by PDG
\cite{pdg}: $1230\pm 40$ MeV and $\sim 400$ MeV. Historically,  
there exists a
respectable old prediction by Weinberg \cite{Wei}, $m(a_1)\simeq  
\sqrt{2}\;m(
\rho ),$ which places the mass of the $a_1$ around 1090 MeV, in apparent
disagreement with experiment. It is reasonable to think that the  
discrepancy
between Weinberg's prediction and the currently adopted value can  
be explained
if one includes possible contributions from the radial and orbital  
excitations
of $\rho $ and $a_1$ to the spectral sum rules used for the  
derivation of the
$a_1$ mass. These excited states can, in fact, possess  
nonnegligible matrix
elements to the vacuum through the vector and axial-vector currents, as 
indicated by the appreciable rate of the $\rho ^{'}\rightarrow  
e^{+}e^{-}$
decay \cite{pdg}. An alternative derivation of the $\rho $-$a_1$  
mass relation
using the both Weinberg's and KSFR \cite{KSFR} sum rules done by Li  
\cite{Li}
leads to $m(a_1)=1.36\pm 0.23$ GeV, in better agreement with  
experiment. The
value of the $a_1$ mass provided by QCD sum rules is \cite{RRY}  
$1150\pm 40$
MeV. Recently Oneda {\it et al.} put forward theoretical arguments which 
suggest a simultaneous mass degeneracy of the axial-vector and  
pseudovector
nonets in the isovector and isodoublet channels \cite{OST}. The  
mass relations
obtained in \cite{OST} in the algebraic approach to QCD developed  
in ref.
\cite{OT},
\bqry
m^2(K^\ast ) & - & m^2(\rho  
)\;=\;m^2(K_{1A})\;-\;m^2(a_1)\;=\;m^2(K_{1B})\;-\;
m^2(b_1), \NL
m(K_{1A}) & = &  m(K_{1B}),
\eqry
lead to $m(a_1)=m(b_1).$ Since, according to the recent PDG  
analysis, $m(b_1)=
1231\pm 10$ MeV, degeneracy of the $a_1$ and $b_1$ meson masses  
seems to be a
real possibility. \\
2) The $q\bar{q}$ model predicts a nonet that includes two  
isoscalar 1 $^3P_1$
states with masses below $\sim $ 1.6 GeV. Three ``good'' $1^{++}$  
objects are
known, the $f_1(1285),\;f_1(1420)$ and $f_1(1510),$ one more than  
expected.
Thus, one of the three is a non-$q\bar{q}$ meson, and the  
$f_1(1420)$ is the
best non-$q\bar{q}$ candidate, according to ref. \cite{Cald}. In  
this case, it
may be a multiquark state in the form of a $K\bar{K}\pi $ bound state 
(``molecule'') \cite{Long}, or a $K\bar{K}^\ast $ deuteron-like state 
(``deuson'') \cite{Torn91}. On the other hand, Aihara {\it et al.}  
\cite{Aih}
have argued that, assuming that both the $f_1(1285)$ and  
$f_1(1420)$ belong
to the same nonet and using several additional hypotheses, the  
octet-singlet
mixing angle obtained is compatible with the $f_1(1420)$ being  
mostly $s\bar{
s}$ and the $f_1(1285)$ being mostly  
$(u\bar{u}+d\bar{d})/\sqrt{2},$ although
both require large admixtures of other $q\bar{q}$ components.
 \\   \\
3. Pseudovector meson nonet.\hfil\break
Experimental information on the $h_1$ and $h_1^{'}$ mesons is rather 
restricted. A very wide $\rho \pi $ resonance with $I=0,$  
$J^{PC}=1^{+-}$ has
been seen in three experiments as having mass and width $1170\pm  
20$ MeV and
$360\pm 40$ MeV, respectively, which is identified with the  
dominantly $(u\bar{
u}+d\bar{d})/\sqrt{2}$ meson $h_1.$ Its $s\bar{s}$ partner $h_1^{'}$ is
expected to decay dominantly into $K\bar{K}\pi .$ So far, in a single 
experiment by LASS, a candidate has been observed in $K^{-}p\rightarrow
K_{S}K^{\pm }\pi ^{\mp }\Lambda $ with mass and width $1380\pm 20$  
MeV and $80
\pm 30$ MeV, respectively, decaying into $(K^\ast  
\bar{K}+\bar{K^\ast }K)$
\cite{A}. More recently, the Crystal Barrel collaboration has studied the
reaction $\bar{p}p\rightarrow K_{L}K_{S}\pi ^0\pi ^0 (\rightarrow  
8\gamma \;+$
missing energy and momentum) \cite{Felix}. Its final state is  
dominated by the
strange resonances $K_1(1400),$ $K_1(1270),$ $K_0^\ast (1430)$ and  
$K^\ast (
892).$ However, albeit with a small intensity, the best fit  
requires also a
$1^{+}$ state at 1385 MeV with width 200 MeV, decaying into $K^\ast  
K.$ The
$K_{S}K_{L}\pi ^0$ Dalitz plot for a mass window of $\pm 50$ MeV  
around 1380
MeV shows a clearly destructive interference between the two  
$K^\ast $ bands,
which is consistent with an $I=0,$ $J^{PC}=1^{+-}$ state.
 \\   \\
4. Tensor meson nonet.\hfil\break
The two 1 $^3P_2$ $q\bar{q}$ states are likely the well-known  
$f_2(1270)$ and
$f_2^{'}(1525)$ currently adopted by PDG, although the observation by 
Breakstone \cite{Break} of the $f_2(1270)$ production by gluon  
fusion could
indicate that it has a glueball component. At least five more  
$J^{PC}=2^{++}$
states have to be considered: the $f_2(1520),\;f_2(1810),
\;f_2(2010),\;f_2(2300)$ and $f_2(2340).$ Of these, the $f_2(1810)$  
is likely
to be the 2 $^3P_2$, and the three $f_2$'s above 2 GeV could  
possibly be the 2
$^3P_2$ $s\bar{s}$, 1 $^3F_2$ $s\bar{s},$ and 3 $^3P_2$ $s\bar{s},$  
but a
gluonium interpretation of one of the three is not excluded. The  
remaining
$f_2(1520)$ was seen in 1989 by the ASTERIX collaboration  
\cite{Ast} as a $2^{
++}$ resonance in $p\bar{p}$ $P$-wave annihilation at 1565 MeV in  
the $\pi ^{+
}\pi ^{-}\pi ^0$ final state. Its mass is better determined in the  
$3\pi ^0$
mode by the Crystal Barrel collaboration \cite{CB} to be 1515 MeV, in 
agreement with that seen previously \cite{BG}. It has no place in a 
$q\bar{q}$ scheme mainly because all nearby $q\bar{q}$ states are  
already
occupied. Dover \cite{Dover} has suggested that it is a ``quasinuclear'' 
$N\bar{N}$ bound state, and T\"{o}rnqvist \cite{Torn91} that it is a 
deuteron-like $(\omega \omega +\rho \rho )/\sqrt{2}$ ``deuson'' state.
 \\   \\
5. Let us also discuss the $I=1/2$ 1 $^3P_1$ and 1 $^1P_1$ mesons,  
$K_1(1270)$
and $K_1(1400),$ with masses $1273\pm 7$ MeV and $1402\pm 7$ MeV,
respectively \cite{pdg}. It has been known that their decay satisfies a 
dynamical selection rule
$$\Gamma \left( K_1(1270)\rightarrow K\rho \right) >> \Gamma \left(  
K_1(1270)
\rightarrow K^\ast \pi \right) ,$$
$$\Gamma \left( K_1(1400)\rightarrow K^\ast \pi \right) >> \Gamma  
\left( K_1(
1400)\rightarrow K\rho \right) ,$$
which prompted experimentalists to suspect large mixing (with a  
mixing angle
close to $45^o)$ between the $I=1/2$ members of the axial-vector and 
pseudovector nonets, $K_{1A}$ and $K_{1B},$ respectively, leading to the 
physical $K_1$ and $K_1^{'}$ states \cite{K1}. Carnegie {\it et al.}
\cite{Car} obtained the mixing angle $\theta _K=(41\pm 4)^o$ as the 
optimum fit to the data as of 1977. In a recent paper by Blundell  
{\it et al.}
\cite{BGP}, who have calculated strong OZI-allowed decays in the
pseudoscalar emission model and the flux-tube breaking model, the  
$K_{1A}$-$K_{
1B}$ mixing angle obtained is $\simeq 45^o.$ Theoretically, in the exact 
$SU(3)$ limit the $K_{1A}$ and $K_{1B}$ states do not mix,  
similarly to their
$I=1$ counterparts $a_1$ and $b_1.$ For the $s$-quark mass greater  
than the
$u$- and $d$-quark masses, $SU(3)$ is broken and these states mix  
to give the
physical $K_1$ and $K_1^{'}.$ If the $K_{1A}$ and $K_{1B}$ are  
degenerate
before mixing, the mixing angle will always be $\theta _K=45^o$  
\cite{CR,Lip}.
However, as pointed out by Suzuki \cite{Suz}, the recent data of the 
TPC/Two-Gamma collaboration on $K\pi \pi $ production in $\tau $-decay 
appear to contradict this simple picture: if $\theta _K=45^o,$  
production of
the $K_1(1270)$ and $K_1(1400)$ would be one-to-one up to the kinematic 
corrections, since in the $SU(3)$ limit only the linear combination  
$\left( K_
1(1270)+K_1(1400)\right) /\sqrt{2}$ would have the right quantum  
number to be
produced there. After the phase-space correction, the $K_1(1270
)$ production would be favored over the $K_1(1400)$ one by nearly a  
factor of
2. Actually, $K\pi \pi $ production is dominated by the  
$K_1(1400),$ with
little evidence for the $K_1(1270).$ As found by Suzuki \cite{Suz},  
the $K_1(
1270)/K_1(1400)$ production ratio observed favors $\theta _K\approx  
33^o,$
although some $SU(3)$ breaking effects are needed to obtain good  
qualitative
agreement between the theory and experiment. \\

Since the experimentally established isodoublet states of the scalar
and tensor meson nonets, $K_0^\ast $ and $K_2^\ast ,$ are mass  
degenerate,
$1429\pm 6$ MeV and 1429 MeV, respectively, and different models  
(like those
considered in refs. \cite{Klempt,Dmitra,invited}) lead to the $q\bar{q}$
assignment for the scalar nonet which includes both the $a_0(1320)$  
and $f_0(
1525)$ mesons which are mass degenerate with the corresponding  
tensor mesons
$a_2(1320)$ and $f_2^{'}(1525),$ the question naturally suggests  
itself as to
whether the scalar and tensor nonets are intrinsically mass
degenerate\footnote{In the scenario suggested in refs.  
\cite{Klempt,Dmitra},
due to instanton effects, the mass of the $f_0^{'}$ meson is  
shifted down to
$\sim 1$ GeV, as compared to the mass $\approx 1275$ MeV of its tensor 
``partner'' $f_2.$} \cite{enigmas}. Similar questions may be asked  
regarding
the mass degeneracy of the axial-vector and pseudovector nonets in  
the $I=1$
and $I=1/2$ channels. If this mass degeneracy of two pairs of  
nonets, $(^3P_0-
^3P_2)$ and $(^3P_1-^1P_1),$ is actually the case, it should be  
reproduced in a
simple phenomenological model of QCD, e.g., in a nonrelativistic  
constituent
quark model. The purpose of this work is to apply the latter model for 
$P$-wave meson spectroscopy in order to establish whether mass  
degeneracy of
the two pairs of nonets discussed above actually occurs.

\section{Nonrelativistic constituent quark model}
In the constituent quark model, conventional mesons are bound  
states of a spin
1/2 quark and spin 1/2 antiquark bound by a phenomenological  
potential which
has some basis in QCD \cite{LSG}. The quark and antiquark spins  
combine to
give a total spin 0 or 1 which is coupled to the orbital angular  
momentum $L.$
This leads to meson parity and charge conjugation given by  
$P=(-1)^{L+1}$ and
$C=(-1)^{L+S},$ respectively. One typically assumes that the  
$q\bar{q}$ wave
function is a solution of a nonrelativistic Schr\"{o}dinger  
equation with the
generalized Breit-Fermi Hamiltonian\footnote{The most widely used  
potential
models are the relativized model of Godfrey and Isgur \cite{GI} for the 
$q\bar{q}$ mesons, and Capstick and Isgur \cite{CI} for the $qqq$  
baryons.
These models differ from the nonrelativistic quark potential model  
only in
relatively minor ways, such as the use of $H_{kin}=\sqrt{m_1^2+{\bf  
p}_1^2}+
\sqrt{m_2^2+{\bf p}_2^2}$ in place of that given in (8), the  
retention of the
$m/E$ factors in the matrix elements, and the introduction of coordinate
smearing in the singular terms such as $\delta ({\bf r}).$}, $H_{BF},$
\beq
H_{BF}\;\psi _n({\bf r})\equiv \left( H_{kin}+V({\bf p},{\bf  
r})\right) \psi _
n({\bf r})=E_n\psi _n({\bf r}),
\eeq
where $H_{kin}=m_1+m_2+{\bf p}^2/2\mu -(1/m_1^3+1/m_2^3){\bf  
p}^4/8,$ $\mu =m_
1m_2/(m_1+m_2),$ $m_1$ and $m_2$ are the constituent quark masses, and to
first order in $(v/c)^2={\bf p}^2c^2/E^2\simeq {\bf p}^2/m^2c^2,$  
$V({\bf p},
{\bf r})$ reduces to the standard nonrelativistic result,
\beq
V({\bf p},{\bf r})\simeq V(r)+V_{SS}+V_{LS}+V_T,
\eeq
with $V(r)=V_V(r)+V_S(r)$ being the confining potential which  
consists of a
vector and a scalar contribution, and $V_{SS},V_{LS}$ and $V_T$ the  
spin-spin,
spin-orbit and tensor terms, respectively, given by
\cite{LSG}
\beq
V_{SS}=\frac{2}{3m_1m_2}\;{\bf s}_1\cdot {\bf s}_2\;\triangle V_V(r),
\eeq
$$V_{LS}=\frac{1}{4m_1^2m_2^2}\frac{1}{r}\left( \left\{  
[(m_1+m_2)^2+2m_1m_2]\;
{\bf L}\cdot {\bf S}_{+}+(m_2^2-m_1^2)\;{\bf L}\cdot {\bf S}_{-}\right\}
\frac{dV_V(r)}{dr}\right. $$
\beq
\left. -\;[(m_1^2+m_2^2)\;{\bf L}\cdot {\bf  
S}_{+}+(m_2^2-m_1^2)\;{\bf L}\cdot
{\bf S}_{-}]\;\frac{dV_S(r)}{dr}\right) ,
\eeq
\beq
V_T=\frac{1}{12m_1m_2}\left(  
\frac{1}{r}\frac{dV_V(r)}{dr}-\frac{d^2V_V(r)}{
dr^2}\right) S_{12}.
\eeq
Here ${\bf S}_{+}\equiv {\bf s}_1+{\bf s}_2,$ ${\bf S}_{-}\equiv  
{\bf s}_1-
{\bf s}_2,$ and
\beq
S_{12}\equiv 3\left( \frac{({\bf s}_1\cdot {\bf r})({\bf s}_2\cdot  
{\bf r})}{
r^2}-\frac{1}{3}{\bf s}_1\cdot {\bf s}_2\right).
\eeq
For constituents with spin $s_1=s_2=1/2,$ $S_{12}$ may be rewritten  
in the form
\beq
S_{12}=2\left( 3\frac{({\bf S}\cdot {\bf r})^2}{r^2}-{\bf  
S}^2\right),\;\;\;
{\bf S}={\bf S}_{+}\equiv {\bf s}_1+{\bf s}_2.
\eeq
Since $(m_1+m_2)^2+2m_1m_2=6m_1m_2+(m_2-m_1)^2,$  
$m_1^2+m_2^2=2m_1m_2+(m_2-m_
1)^2,$ the expression for $V_{LS},$ Eq. (11), may be rewritten as  
follows,
$$V_{LS}=\frac{1}{2m_1m_2}\frac{1}{r}\left[ \left( 3\frac{dV_V(r)}{dr}-
\frac{dV_S(r)}{dr}\right) +  
\frac{(m_2-m_1)^2}{2m_1m_2}\left(\frac{dV_V(r)}{d
r}-\frac{dV_S(r)}{dr}\right) \right] {\bf L}\cdot {\bf S}_{+}$$
\beq
+\frac{m_2^2-m_1^2}{4m_1^2m_2^2}\;\frac{1}{r}\left(  
\frac{dV_V(r)}{dr}-\frac{
dV_S(r)}{dr}\right) {\bf L}\cdot {\bf S}_{-}\equiv V_{LS}^{+}+V_{LS}^{-}.
\eeq
Since two terms corresponding to the derivatives of the potentials  
with respect
to $r$ are of the same order of magnitude, the above expression for 
$V_{LS}^{+}$ may be rewritten as
\beq
V_{LS}^{+}=\frac{1}{2m_1m_2}\frac{1}{r}\left(  
3\frac{dV_V(r)}{dr}-\frac{dV_
S(r)}{dr}\right) {\bf L}\cdot {\bf S}\left[  
1+\frac{(m_2-m_1)^2}{2m_1m_2}\;O(
1)\right] .
\eeq

\subsection{$S$-wave spectroscopy}
Let us first apply the Breit-Fermi Hamiltonian to the $S$-wave  
which consists
of the two, $^1S_0$ $J^{PC}=0^{-+}$ pseudoscalar and $^3S_1$  
$1^{--}$ vector,
meson nonets. We shall consider only the $I=1$ and $I=1/2$ mesons  
which are
pure $n\bar{n}$ and $(n\bar{s},\;s\bar{n})$ states, respectively.  
Since the
expectation values of the spin-orbit and tensor terms vanish for  
$L=0$ or $S=
0$ states \cite{LSG}, the mass of a $q\bar{q}$ state with $L=0$ is  
given by
\beq
M=m_1+m_2+E+K\;\frac{{\bf s}_1\cdot {\bf s}_2}{m_1m_2},
\eeq
where $E$ is the nonrelativistic binding energy. As shown below,  
the sum of
just the constituent quark masses and the quark-quark hyperfine  
interaction
term describes the masses of the $S$-wave mesons extremely well.  
Moreover, Eq.
(17) with no $E$ is consistent with the empirical mass squared splitting 
$\triangle M^2\equiv M^2(^3S_1)-M^2(^1S_0)\approx $ const for all the 
corresponding mesons composed by the $n,s$ and $c$ quarks, aside from 
charmonia. Physically, these observations mean that $E$ is small  
compared to
$m_1$ and $m_2$ or approximately constant over all of these states,  
and so may
be absorbed in the definition of the latter. For the higher $L$  
nonets, $E$
decreases with the increasing quark masses, according to the  
Feynman-Hellmann
theorem; it therefore may be absorbed into the constituemt quark  
mass defined
for every  $L.$

Since ${\bf s}_1\cdot {\bf s}_2=-3/4$ for spin-0 mesons and $+1/4$  
for spin-1
mesons, one has the four relations (in the following, $\pi $ stands  
for the
mass of the $\pi $ meson, etc., $n$ and $s$ for the masses of the
non-strange and strange quarks, respectively, and we assume $SU(2)$  
flavor
symmetry, $m_u=m_d=n$),
\bqry
\pi & = & 2\;n\;-\;\frac{3}{4}\;\frac{\Lambda }{n^2}, \\
\rho & = & 2\;n\;+\;\frac{1}{4}\;\frac{\Lambda }{n^2}, \\
K & = & n+s-\frac{3}{4}\frac{\Lambda }{ns}, \\
K^\ast \! & = & n+s+\frac{1}{4}\frac{\Lambda }{ns}.
\eqry
It then follows from these relations that
\bqry
n & = & \frac{\pi +3\rho }{8}, \\
s & = & \frac{2K+6K^\ast -\pi -3\rho }{8}, \\
\frac{\Lambda }{n^2} & = & \frac{\rho -\pi }{2}, \\
\frac{\Lambda }{ns} & = & \frac{K^\ast -K}{2}.
\eqry
By expressing the ratio $n/s$ in two different ways, directly from  
(22),(23)
and dividing (25) by (24), one obtains the relation
\beq
\frac{n}{s}=\frac{\pi +3\rho }{2K+6K^\ast -\pi -3\rho  
}=\frac{K^\ast -K}{\rho
-\pi }.
\eeq
For the physical values of $\pi ,\rho ,K$ and $K^\ast $ (in MeV),  
138, 769,
495 and 892, respectively, the above relation reads $0.629=0.627,$ i.e., 
the result is consistent within
the accuracy provided by the assumption of exact $SU(2)$ flavor  
symmetry. The
values of $n,s$ and $K$ provided by (22)-(25) are $n=306$ MeV,  
$s=487$ MeV,
$K=0.0592$ GeV$^3=(390\;{\rm MeV})^3.$ The values of the meson  
masses, as
calculated from (18)-(21), are (in MeV) $\pi =137.8,$ $\rho =770.0,$ 
$K=495.0,$ $K^\ast =892.3.$ The relation (17) may also be applied
successfully to the $^3S_1$ $I=0$ mesons too, assuming that they  
are pure $n
\bar{n}$ and $s\bar{s}$ states. In this case, as follows from (19),  
$\omega =
\rho =770$ MeV, and $\phi =2s+K/(4s^2)=1036\;{\rm MeV}.$ Both  
numbers are
within $1.5$\% of the physical values 782 and 1019 MeV, respectively. 

Let us note that, although Eq. (18) contains no information on  
chiral symmetry,
one may deal with the chiral limit $\pi =0$ by the introduction of the so
called ``dynamical'' quark mass \cite{ETS}, $m_{dyn},$ defined as  
the solution
of $2m_{dyn}-3K/(4m_{dyn}^2)=0.$ Although this does not restore chiral 
symmetry, it does incorporate the masslessness of the
pion, in accord with common understanding of the latter as the  
Nambu-Goldstone
boson of broken chiral symmetry, as well as calculating the chiral limit 
values of $\rho $ and $K^\ast ,$ in agreement with other models  
\cite{other}.

\section{$P$-wave spectroscopy}
We now wish to apply the Breit-Fermi Hamiltonian to the $P$-wave  
mesons. By
calculating the expectation values of different terms of the Hamiltonian 
defined in Eqs. (10),(14),(15), taking into account the  
corresponding matrix
elements $\langle {\bf L}\cdot {\bf S}\rangle $ and $S_{12}$  
\cite{LSG}, one
obtains the relations \cite{BGP}
\bqryn
M(^3P_0) & = & M_0+\frac{1}{4}\langle V_{SS}\rangle -2\langle V_{LS}^{+}
\rangle + \langle V_T\rangle , \\
M(^3P_2) & = & M_0+\frac{1}{4}\langle V_{SS}\rangle +\langle  
V_{LS}^{+}\rangle
+ \frac{1}{10}\langle V_T\rangle , \\
M(a_1) & = & M_0+\frac{1}{4}\langle V_{SS}\rangle -\langle  
V_{LS}^{+}\rangle -
\frac{1}{2}\langle V_T\rangle , \\
M(b_1) & = & M_0-\frac{3}{4}\langle V_{SS}\rangle ,
\eqryn
$$\left( \begin{array}{c}
M(K_1) \\ M(K_1^{'}) \end{array} \right) =\left( \begin{array}{cc}
M_0+\frac{1}{4}\langle V_{SS}\rangle -\langle V_{LS}^{+}\rangle  
-\frac{1}{2}
\langle V_T\rangle  & \sqrt{2}\langle V_{LS}^{-}\rangle \\
\sqrt{2}\langle V_{LS}^{-}\rangle  & M_0-\frac{3}{4}\langle  
V_{SS}\rangle
\end{array} \right) \left( \begin{array}{c}
K_{1A} \\ K_{1B} \end{array} \right) ,$$
where $M_0$ stands for the sum of the constituent quark masses in  
either case.
The $V_{LS}^{-}$ term acts only on the $I=1/2$ singlet and triplet  
states
giving rise to the spin-orbit mixing between these states\footnote{The 
spin-orbit $^3P_1-^1P_1$ mixing is a property of the model we are
considering; the possibility that another mechanism is responsible  
for this
mixing, such as mixing via common decay channels \cite{Lip} should  
not be ruled
out, but is not included here.}, and is responsible for the  
physical masses of
the $K_1$ and $K_1^{'}.$
Let us assume, for simplicity, that $$\sqrt{2}\langle  
V_{LS}^{-}\rangle (K_{
1B})\simeq -\sqrt{2}\langle V_{LS}^{-}\rangle (K_{1A})\equiv \Delta  
.$$ The
masses of the $K_{1A},\;K_{1B}$ are then determined by relations  
similar to
those for the $a_1,\;b_1$ above, and $K_1\simeq  
K_{1A}+\Delta,\;K_1^{'}\simeq
K_{1B}-\Delta ,$ or\footnote{Actually, as follows from Eq. (45)  
below, $$\frac{
K_1-K_{1A}}{K_{1B}-K_1^{'}}=\frac{K_1^{'}+K_{1B}}{K_1+K_{1A}}\simeq  
\frac{2K_{
1B}}{2K_{1A}}\simeq 1,$$ since the deviations  
$K_1-K_{1A},\;K_{1B}-K_1^{'}$ are
small compared to $K_{1A},\;K_{1B},$ and the mixing angle is $\sim  
45^o.$}
\beq
\Delta \simeq K_1-K_{1A}\simeq K_{1B}-K_1^{'}.
\eeq
We consider, therefore, the following formulas for the masses of  
all eight $I=
1,1/2$ $P$-wave mesons, $b_1,a_0,a_1,a_2,K_{1B},K_0^\ast  
,K_{1A},K_2^\ast :$
\bqry
M(^1P_1) & = & m_1+m_2-\frac{3}{4}\frac{a}{m_1m_2}, \\
M(^3P_0) & = &  
m_1+m_2+\frac{1}{4}\frac{a}{m_1m_2}-\frac{2b}{m_1m_2}+\frac{c}{
m_1m_2}, \\
M(^3P_1) & = &  
m_1+m_2+\frac{1}{4}\frac{a}{m_1m_2}-\frac{b}{m_1m_2}-\frac{c}{
2m_1m_2}, \\
M(^3P_2) & = &  
m_1+m_2+\frac{1}{4}\frac{a}{m_1m_2}+\frac{b}{m_1m_2}+\frac{c}{
10m_1m_2},
\eqry
where $a,b$ and $c$ are related to the matrix elements of $V_{SS},$  
$V_{LS}$
and $V_T$ (see Eqs. (10),(12),(16)), and assumed to be the same for  
all of the
$P$-wave states. In the above expressions, the nonrelativistic  
binding energies
are absorbed in the constituent quark masses, as discussed above.  
The same
constituent quark masses appear also in the denominators of the  
hyperfine
interaction terms in Eqs. (12)-(15) below, similar to $S$-wave  
spectroscopy
considered in a previous section. Since this is usually done only  
for the
lowest $S$-wave states, we briefly review the precedent and  
argument for the
generality of these forms.

It was shown in \cite{scalar} that a pure scalar potential  
contributes to the
effective constituent quark mass. Bag models suggest that the kinetic 
energy also contributes to the effective constituent quark mass in  
the case of
no potential \cite{bag}. These results were generalized further by  
Cohen and
Lipkin \cite{CL} who have shown that both the kinetic and potential  
energy are
included in the effective mass parameter which appears also in the  
denominators
of the hyperfine interaction terms in the case of a scalar  
confining potential.
The analyses of experimental data suggest that the non-strange and  
strange
quarks are mainly subject to scalar part of the confining potential  
(whereas
charmed and bottom quarks are more dominantly affected by  
Coulomb-like vector
part) \cite{LSG}. Moreover, the generality of the arguments by Cohen and 
Lipkin \cite{CL} allows one to apply them to any partial wave.  
Therefore, the
constituent quark masses can be defined for any partial wave, through 
relations of the form (28)-(31); in this case they vary with the  
energies of
the corresponding mass levels. Such an energy dependence of the  
constituent
quark masses was considered in refs. \cite{GY,BB}. Also, a  
QCD-based mechanism
which generates dynamical quark mass growing with $L$ in a  
Regge-like manner
was considered by Simonov \cite{Sim}.

The correction to $V_{LS}^{+}$ in the formula (16), due to the  
difference in
the masses of the $n$ and $s$ quarks, is ignored. Indeed, these  
masses, as
calculated from (28)-(31), are
\beq
n=\frac{3b_1+a_0+3a_1+5a_2}{24},
\eeq
\beq
s=\frac{6K_{1B}+2K_0^\ast +6K_{1A}+10K_2^\ast -3b_1-a_0-3a_1-5a_2}{24}.
\eeq
With the physical values of the meson masses (in GeV), $a_1\simeq  
b_1\cong
1.23,$ $a_0\simeq a_2\cong 1.32,$ $K_{1A}\simeq K_{1B}\cong 1.34,$  
$K_0^\ast
\simeq K_2^\ast \cong 1.43,$ the above relations give
\beq
n\simeq 640\;{\rm MeV,}\;\;\;s\simeq 740\;{\rm MeV,}
\eeq
so that the abovementioned correction, according to (16), is $\sim  
100^2/(2
\cdot 640\cdot 740)\simeq 1$\%, i.e., comparable to isospin  
breaking on the
scale considered here, and so completely negligible. It follows  
from (28)-(31)
that
\bqry
\frac{9a}{m_1m_2} & = & M(^3P_0)+3M(^3P_1)+5M(^3P_2)-9M(^1P_1), \\
\frac{12b}{m_1m_2} & = & 5M(^3P_2)-3M(^3P_1)-2M(^3P_0), \\
\frac{18c}{5m_1m_2} & = & 2M(^3P_0)+M(^3P_2)-3M(^3P_1).
\eqry
By expressing the ratio $n/s$ in four different ways, viz.,  
directly from
(32),(33) and dividing the expressions (35)-(37) for the $I=1/2$  
and $I=1$
mesons by each other, similarly to the case of the $S$-wave mesons  
considered
above, one obtains the three relations,
\beq
\frac{3a_1+3b_1+a_0+5a_2}{6K_{1A}+6K_{1B}+2K_0^\ast +10K_2^\ast  
-3a_1-3b_1-
a_0-5a_2}=\frac{K_0^\ast +3K_{1A}+5K_2^\ast  
-9K_{1B}}{a_0+3a_1+5a_2-9b_1},
\eeq
\beq
\frac{K_0^\ast +3K_{1A}+5K_2^\ast  
-9K_{1B}}{a_0+3a_1+5a_2-9b_1}=\frac{5K_2^
\ast -3K_{1A}-2K_0^\ast }{5a_2-3a_1-2a_0},
\eeq
\beq
\frac{5K_2^\ast -3K_{1A}-2K_0^\ast  
}{5a_2-3a_1-2a_0}=\frac{2K_0^\ast +K_2^
\ast -3K_{1A}}{2a_0+a_2-3a_1}.
\eeq
First consider Eq. (40) which may be rewritten, by a simple algebra, as
\beq
(K_2^\ast -K_0^\ast )(a_2-a_1)=(K_2^\ast -K_{1A})(a_2-a_0).
\eeq
Since $K_2^\ast \cong K_0^\ast \approx 1430$ MeV, it then follows  
from (41)
that either $K_2^\ast \cong K_0^\ast \cong K_{1A},$ or $a_0\cong  
a_2.$ The
first possibility should be discarded as unphysical, since it  
leads, through
the relations (36),(37) applied to the $I=1/2$ mesons, to $b=c=0,$  
which would
in turn, from the same relations for the $I=1$ mesons, imply  
$a_0\cong a_1
\cong a_2,$ in apparent contradiction with experimental data on the  
masses of
the $a_1$ and $a_2$ mesons. The physical case corresponds, therefore, to
\beq
a_0\cong a_2,
\eeq
i.e., the mass degeneracy of the scalar and tensor meson nonets in  
the $I=1/2$
channel, $K_0^\ast \cong K_2^\ast ,$ implies a similar degeneracy  
also in the
$I=1$ channel. Note that this relation is a general feature of the
nonrelativistic quark model for the $P$-wave mesons we are  
considering here.
Even in the presence of an extra term in (28)-(31) corresponding to  
the quark
binding energy which we have ignored by absorbing into the  
constituent masses,
Eqs. (36) and (37) will remain the same and again lead, through  
(40), to the
relation (42).

With $K_0^\ast =K_2^\ast $ and $a_0=a_2,$ Eqs. (38) and (39) may be  
rewritten
as
\beq
(a_0-a_1+K_0^\ast -K_{1A})(a_1+b_1+2a_0)=2(K_0^\ast  
-K_{1A})(K_{1A}+K_{1B}+
2K_0^\ast ),
\eeq
\beq
(K_{1A}-K_{1B})(a_0-a_1)=(K_0^\ast - K_{1A})(a_1-b_1).
\eeq
One now has to determine the values of $a_1,$ $K_{1A}$ and $K_{1B}.$ The 
remaining equation is obtained from the mixing of the $K_{1A}$ and  
$K_{1B}$
states which results in the physical $K_1$ and $K_1^{'}$ mesons;  
independent
of the mixing angle,
\beq
K_{1A}^2+K_{1B}^2=K_1^2+K_1^{'2}.
\eeq
One sees that, as follows from (44), the mass degeneracy of the  
$^3P_1$ and
$^1P_1$ nonets in the $I=1/2$ channel, $K_{1A}=K_{1B},$ implies similar
degeneracy in the $I=1$ channel too, $a_1=b_1,$ and vice versa, so  
that the
model we are considering provides the consistent possibility
\beq
a_1=b_1,\;\;\;K_{1A}=K_{1B}.
\eeq
We now check how this possibility agrees with experimental data on  
the meson
masses. It follows from (45) and $K_1=1273\pm 7$ MeV,  
$K_1^{'}=1402\pm 7$ MeV
that in this case
\beq
K_{1A}=K_{1B}=1339\pm 7\;{\rm MeV}.
\eeq
With $a_1=b_1,$ $K_{1A}=K_{1B},$ Eq. (43) reduces to
\beq
a_0^2-a_1^2+(a_0+a_1)(K_0^\ast -K_{1A})=2(K_0^{\ast 2}-K_{1A}^2),
\eeq
which for $a_0=a_2=1318$ MeV, $K_0^\ast =1429$ MeV and $K_{1A},$ $K_{1B}$
given in (47) has the solution
\beq
a_1=b_1=1211\pm 8\;{\rm MeV},
\eeq
which is only a 2-standard deviation inconsistency with the  
experimentally
established $b_1$ meson mass $1231\pm 10$ MeV. We also consider another 
solution of (43)-(45) determined by adjusting $b_1$ to the  
experimental value
$1231$ MeV. It then follows that in this case the solution to  
(43)-(45) is
\beq
a_1=1191\;{\rm MeV},\;\;\;K_{1A}=1322\;{\rm  
MeV},\;\;\;K_{1B}=1356\;{\rm MeV},
\eeq
with small deviations from these values for possible deviations in  
the input
parameters; e.g., with (in MeV) $b_1=1231\pm 10,$ the actual  
solution is $a_1=
1191\mp 10,$ $K_{1A}=1322\mp 9,$ $K_{1B}=1356\pm 9,$ or with  
$K_1=1273\pm 7,$
$K_1^{'}=1402\pm 7,$ the solution is $a_1=1191\pm 17,$  
$K_{1A}=1322\pm 14,$ and
$K_{1B}$ remains the same. For the solution (50), we observe the sum
rule
\beq
K_{1A}^2-a_1^2=0.329\;{\rm GeV}^2\simeq K_{1B}^2-b_1^2=0.323\;{\rm  
GeV}^2,
\eeq
which is accurate to 2\% and also holds for deviations from (50) due to 
uncertanties in the input parameters. Relations of the type (51) may be 
anticipated on the basis of the formulas $$K^{\ast 2}-\rho  
^2=K^2-\pi ^2,\;\;
\;K_2^{\ast 2}-a_2^2=K^2-\pi ^2,\;\;\;{\rm etc.,}$$ provided by  
either the
algebraic approach to QCD \cite{OT} or phenomenological formulas  
$$m_1^2=2Bn+
C,\;\;\;m_{1/2}^2=B(n+s)+C$$ (where $B$ is related to the quark  
condensate,
and $C$ is a constant within a given meson nonet) motivated by the  
linear mass
spectrum of a nonet and the collinearity of Regge trajectories of the 
corresponding $I=1$ and $I=1/2$ states, as discussed in ref.  
\cite{linear}.
Note that (51) agrees with the second of the three relations (7).

Thus, the nonrelativistic constituent quark model we are  
considering provides
two possibilities for the mass spectra of the axial-vector and  
pseudovector
meson nonets:

1) $a_1=b_1\simeq 1210\;{\rm MeV},\;\;\;K_{1A}=K_{1B}\simeq  
1340\;{\rm MeV},$

2) $a_1\neq b_1,\;K_{1A}\neq  
K_{1B},\;\;\;K_{1A}^2\;-\;a_1^2\;\simeq \;K_{1B
}^2\;-\;b_1^2.$ \\
The second case is obviously favored by current experimental data  
on $K\pi
\pi $ production in $\tau $-decay which do not support $\theta _K\approx 
45^o$ and, therefore, mass degeneracy of the $K_{1A}$ and $K_{1B},$ as 
discussed above in the text. In this case, assume that the  
$K_1(1270)$ belongs
to the axial-vector nonet, while the $K(1400)$ belongs to the  
pseudovector
nonet, in accord with the recent suggestion by Suzuki \cite{Suz2},  
on the basis
of the analysis of the $\tau $-decay mode $\tau \rightarrow \nu  
_\tau K_1,$ for
the values (in MeV, as follows from discussion below Eq. (50))  
$K_1=1273\pm 7,$
$K_1^{'}=1402\pm 7,$ $K_{1A}=1322\pm 14,$ $K_{1B}=1356.$ One then  
obtains,
with the help of the formula \cite{Suz}
$$\tan ^2(2\theta _K)=\left(  
\frac{K_1^2(1400)-K_1^2(1270)}{K_{1B}^2-K_{1A}^2}
\right) ^2-1,$$
\beq
\theta _K=(37.3\pm 3.2)^o,
\eeq
in good qualitative agreement with the values $\approx 33^o$  
suggested by
Suzuki \cite{Suz}, and $\simeq 34^o$ found by Godfrey and Isgur  
\cite{GI} in a
relativized quark model.

The parameters of the spin-spin, spin-orbit, and tensor interaction  
in our
model may be calculated from Eqs. (35)-(37) with the meson mass  
values obtained
above. In the isodoublet channel, e.g., one obtains
\bqry
\langle V_{SS}\rangle  & = & \frac{a}{ns}\;\simeq \;37\;{\rm MeV}, \\
\langle V_{LS}^{+}\rangle  & = & \frac{b}{ns}\;\simeq \;27\;{\rm MeV}, \\
\langle V_T\rangle  & = & \frac{c}{ns}\;\simeq \;89\;{\rm MeV}.
\eqry
The expectation value $\langle V_{LS}^{-}\rangle $ may be obtained from 
(27),(50):
$$\sqrt{2}\;\langle V_{LS}^{-}\rangle \simeq  
K_1(1270)-K_{1A}=(1273-1322)\;
{\rm MeV}$$
$$\simeq K_{1B}-K_1(1400)=(1356-1402)\;{\rm MeV}\simeq -47.5\;{\rm  
MeV},$$
and therefore
\beq
\langle V_{LS}^{-}\rangle \simeq -33.5\;{\rm MeV},
\eeq
so that both $\langle V_{LS}^{+}\rangle $ and $\langle  
V_{LS}^{-}\rangle $ are
of the very similar magnitude (but opposite in sign).

Using the obtained values of $\langle V_{LS}^{+}\rangle $ and  
$\langle V_{LS
}^{-}\rangle ,$ along with the values of $n$ and $s$ given in (34),  
in Eqs.
(15),(16), one may establish the following relation among the expectation
values of the derivatives of the potentials:
\beq
\left \langle \frac{1}{r}\frac{dV_S(r)}{dr}\right \rangle \simeq 2.8
\left \langle \frac{1}{r}\frac{dV_V(r)}{dr}\right \rangle .
\eeq
In the case of the QCD-motivated Cornell potential \cite{Corn}
\beq
V(r)=-\frac{4}{3}\frac{\alpha _s}{r}+ar,
\eeq
with a spin structure $V=V_V+V_s,$ $V_V=-\frac{4}{3}\frac{\alpha _s}{r},$
$V_S=ar,$ the relation (57) reduces to
\beq
a\langle r^{-1}\rangle \simeq 3.7\;\alpha _s\langle r^{-3}\rangle .
\eeq
Consider now the ratio \cite{LSG}
\beq
\rho =\frac{M(^3P_2)-M(^3P_1)}{M(^3P_1)-M(^3P_0)}.
\eeq
Since the measured masses of the $K_2^\ast $ and $K_0^\ast $  
coincide (as also
do those of the $a_2$ and $a_0,$ as established in Section 3), the  
value of
this ratio is
\beq
\rho =-1.
\eeq
By equating this value of $\rho $ with that given in \cite{LSG} for  
the Cornell
case,
\beq
\rho =\frac{1}{5}\;\frac{8\alpha _s\langle r^{-3}\rangle  
-\frac{5}{2}a\langle
r^{-1}\rangle}{2\alpha _s\langle r^{-3}\rangle -\frac{1}{4}a\langle  
r^{-1}
\rangle },
\eeq
we obtain
\beq
a\langle r^{-1}\rangle =4.8\;\alpha _s\langle r^{-3}\rangle.
\eeq
Comparison of the relation (63) with (59) shows that the nonrelativistic
constituent quark model considered in this paper is completely  
consistent, at
the 25\% level, with the Cornell potential with the spin structure of a 
vector-scalar mixing type. We consider this to be completely  
satisfactory
agreement, since the expectation values and $\alpha _s$ are all purely 
determined in this region for light quark systems.\footnote{For  
$a=1/(2\pi
\alpha ^{'})\simeq 0.18$ GeV$^2,$ where $\alpha ^{'}\simeq 0.9$  
GeV$^{-2}$ is
the universal Regge slope, it follows from the relation \cite{LSG}
$$\Delta M^2\equiv M^2(^3S_1)-M^2(^1S_0)\simeq \frac{32}{9}\alpha  
_sa\approx
0.56\;{\rm GeV}^2$$ that $\alpha _s\simeq 0.9.$ With these values  
of $a$ and
$\alpha_s,$ and in the approximation $$\langle r^{-3}\rangle \sim \frac{
\langle r^{-1}\rangle }{\langle r\rangle ^2},$$ it then follows  
from (59) that
$\langle r\rangle \simeq 0.9$ fm.}

One may now estimate the masses of the isoscalar mesons of the four  
nonets
assuming that they are pure $s\bar{s}$ states, by the application  
of (28)-(31)
with $m_1=m_2=s;$ it then follows that
\beq
h_1^{'}\simeq f_1^{'}\cong 1435\;{\rm MeV},\;\;\;f_0^{'}\simeq  
f_2^{'}\cong
1525\;{\rm MeV}.
\eeq
Hence, the model we are considering suggests that $1^{++}$  
$s\bar{s}$ state
is the $f_1(1420)$ (with mass $1427\pm 2$ MeV) rather than $f_1(1510)$ 
$(1512\pm 4$ MeV) meson, in accord with the arguments of Aihara  
{\it et al.}
\cite{Aih}. The value 1435 given by (64) is within 4\% of the  
$h_1^{'}$ meson
mass, $1380\pm 20.$ Also, the value 1525 given by (66) agrees with
the experimentally established mass of the $f_2^{'}$ meson,  
$1525\pm 5$ MeV.

At this point we call that the nonrelativistic quark model  
predictions on the
masses of the isoscalar states are reliable for all $P$-wave nonets  
except the
scalar nonet. Indeed, as shown by 't Hooft in his study on the  
$U_A(1)$ problem
\cite{tH}, an expansion of the (euclidian) action around the  
one-instanton
solutions of the gauge fields assuming dominance of the zero modes  
of the
fermion fields leads to an effective $2N_f$-fermion interaction  
$(N_f$ being
the number of fermion flavors) not covered by perturbative gluon  
exchange,
which gives an additional contribution to the ordinary confining
quark-antiquark interaction. As shown in ref. \cite{Munz}, due to its 
point-like nature and specific spin structure, the instanton-induced 
interaction in the formulation of 't Hooft acts on the states with  
spin zero
only. The masses of the other mesons with non-vanishing spin are  
therefore
dominantly determined by the confining interaction, leading to the  
conventional
splitting and an ideal mixing of the $q\bar{q}$ nonets which are well 
reproduced by constituent quark models. The only two nonets whose  
mass spectra
turn out to be affected by an instanton-induced interaction are  
spin zero
pseudoscalar and scalar nonets. Quantitatively, an instanton-induced 
interaction for the scalar mesons is of the same magnitude but  
opposite in
sign to that of the pseudoscalars \cite{Klempt}. It, therefore,  
lowers the
mass of the scalar isosinglet state, in contrast to the case of the 
pseudoscalar isosinglet $(\eta _0)$ state the mass of which is  
pushed up by the
instanton-induced interaction before it mixes with the pseudoscalar  
iso-octet
$(\eta _8)$ state to form the physical $\eta $ and $\eta ^{'}$ states.

Thus, the only nonrelativistic quark model prediction of the masses  
of the
isoscalar states of the scalar nonet which may be trustworthy is  
that on the
mass of the mostly isoscalar octet state (which has a dominantly  
$s\bar{s}$
component\footnote{The scalar meson nonet cannot be ideally mixed,  
as well as
the pseudoscalar one, as seen, e.g., in Eq. (4).}): $f_0\simeq  
1525$ MeV, as
given in (64), in good agreement with the measured mass of the  
$f_0(1500).$
The second isoscalar state of the scalar nonet should be mostly  
$n\bar{n}$ but
also contain a non-negligible $s\bar{s}$ component. Its mass may be  
determined
from Eq. (4) with $f_0=1503\pm 11$ MeV \cite{pdg}: $f_0^{'}=1048\mp  
16$ MeV.

Hence, one of the two, $f_0(980)$ or $f_0(1000),$ may be associated  
with the
remaining isoscalar, which is difficult to decide (that is, on the  
basis of
the constituent quark model we are discussing). However, two  
observations
support the interpretation of the $f_0(980)$ as a $q\bar{q}$ state.  
First, the
$t$-dependence of the $f_0(980)$ and the broad background produced  
in $\pi ^{-
}p\rightarrow \pi ^0\pi ^0n$ differ substantially \cite{Alde}. The  
$f_0(1000)$
is produced in peripheral collisions only, while the $f_0(980)$  
shows a strong
$t$-dependence, as expected for a $q\bar{q}$ state. Second, as  
remarked above,
although the isoscalar mostly $SU(3)$ singlet state should have a  
dominant $n
\bar{n}$ component, its $s\bar{s}$ component should be appreciable.  
The $f_0(
980)$ is seen strongly in $J/\Psi \rightarrow \phi f_0(980),$ but  
at most
weakly in $J/\Psi \rightarrow \omega f_0(980).$ On the basis of  
quark diagrams,
one must conclude that the $f_0(980)$ has a very large $s\bar{s}$  
component;
its decay into $\pi \pi $ with the corresponding branching ratio 78\% 
\cite{pdg} underlines an appreciable $n\bar{n}$ component.

Thus, the constituent quark model discussed in this paper supports  
the $q\bar{
q}$ assignment for the scalar meson nonet (6) found by one of the authors
in a previous paper \cite{invited}. For this assignment, the  
$f_0$-$f_0^{'}$
mixing angle, as calculated with the help of the Gell-Mann--Okubo  
mass formula
for $f_0^{'}=980\pm 10$ MeV \cite{pdg} and $f_0=1525\pm 5$ MeV  
\cite{pdg1},
\beq
\tan ^2\theta _S=\frac{4K_0^{\ast  
2}-a_0^2-3f_0^2}{3f_0^{'2}+a_0^2-4K_0^{\ast
2}},
\eeq
is $$\theta _S=(21.4\pm 1.0)^o,$$ in reasonbly good agreement with the 
value predicted by Ritter {\it et al.} \cite{Klempt1}, $\theta  
_S\approx 25^o,$
for which the partial widths of the $f_0(1500)$ calculated in their  
paper are
$$\pi \pi :\eta \eta :\eta \eta ^{'}:K\bar{K}=1.45:0.32:0.18:0.03,$$ in
excellent agreement with the experimentally observed partial  
widths, Eq. (1).

\section{Concluding remarks}
As we have shown, a nonrelativistic constituent quark model confirms a
simultaneous mass degeneracy of the scalar and tensor nonets in the  
isovector
and isodoublet channels, and suggests a nearly mass-degeneracy of the 
corresponding isoscalar mostly octet states. The mass of the  
remaining $0^{++}$
isoscalar mostly singlet state is probably shifted down to $\sim 1$  
GeV due to
instanton effects, as discussed in refs. \cite{Klempt,Dmitra}, thus  
leaving
two, the $f_0(980)$ and $f_0(1000),$ mesons as candidates for this  
state. Out
of these two, preference should be given to the $f_0(980),$ as  
discussed above
in the text. Let us note that, if one ignores instanton or any  
other effects
which may cause a shift in the mass of the $f_0^{'},$ one would  
arrive at a
$q\bar{q}$ assignment for the scalar nonet which would be nearly mass 
degenerate in all isospin channels (e.g., $f_0(1300)$ in (6) in  
place of $f_0(
980),$ as compared to $f_2(1270)).$ In this case, one would have  
the scalar
nonet almost ideally mixed, just like the tensor one is. Then, as  
shown by
T\"{o}rnqvist \cite{Tor}, flavor symmetry (which should be good in  
the case of
such an ideally mixed nonet) would predict the total width of the  
$a_0(1320)$
(using the experimental $K_0^\ast $ width as normalization) $\Gamma  
>400$ MeV,
and a similar $\sim 400$ MeV width of the $f_0(1525),$ much larger  
than 130
MeV found by GAMS for the $a_0(1320)$ and $\approx 90$ MeV found by  
LASS for
the $f_0(1525).$ Therefore, this case should be considered as unphysical.

Although the possibility of a simultaneous mass degeneracy of the  
axial-vector
and pseudovector nonets in the $I=1$ and $I=1/2$ channels is not  
excluded in
the model considered here, it is disfavored by current experimental data.
By adjusting the mass of the $b_1$ meson to the experimentally  
established
value, the masses of the $a_1,$ $K_{1A}$ and $K_{1B}$ mesons were  
calculated,
leading to the predictions $m(a_1)\simeq 1190$ MeV, and $\theta  
_K\simeq (37\pm
3)^o.$ While the former number naturally interpolates between various 
predictions and current experimental data (e.g., it is at the upper  
limit of
the range $(1150\pm 40)$ MeV established in \cite{RRY}, and at the  
lower limit
of that provided by data, $(1230\pm 40)$ MeV \cite{pdg}), the  
latter one is in
quantitative agreement with the predictions $\theta _K\approx 33^o$  
by Suzuki
\cite{Suz} and $\simeq 34^o$ by Godfrey and Isgur \cite{GI}. The  
results of the
work suggest that the mostly $s\bar{s}$ state of the axial-vector  
nonet should
be associated with the $f_1(1420)$ rather than $f_1(1510)$ meson, which 
supports conclusions of Aihara {\it et al.} \cite{Aih}. We did not  
calculate
the decay widths and branching ratios for this case, since that was  
done in
ref. \cite{Aih}. We wish to give yet another argument in support of this 
prediction. As follows from the chiral theory of mesons initiated by Li 
\cite{Li1}, the mixing angles of both the vector and axial-vector nonets
should coincide \cite{prep1}. The value of the mixing angle of the
axial-vector nonet, as calculated from Eq. (65) for $a_1,\;K_{1A}$  
given in
(50), with deviations due to the
input parameters $K,\;K^{'},$ and $f_1=1427$ MeV, is $$\theta  
_A=(42.4\pm 5.3
)^o,$$ in good agreement with $\theta _V=39.5^o$ of the vector  
meson nonet,
while for $f_1=1512$ MeV and the same $a_1$ and $K_{1A},$ Eq. (65) gives 
$$\theta _A=(54.8\pm 3.4)^o,$$ in apparent disagreement with  
$\theta _V.$

The values of the $a_1$ and $K_{1A}$ masses calculated in this work  
fix the
mass of the $K_{1B}$ to be 1356 MeV. The mass of the isoscalar  
octet state of
the 1 $^1P_1$ nonet is then determined by the Gell-Mann--Okubo  
formula $$h_8^2
=\frac{4K_{1B}^2-b_1^2}{3},$$ $h_8=1395\mp 3$ MeV (for $b_1=1231\pm  
10$ MeV).
Since for the pseudovector nonet Eq. (65) may be rewritten as  
$$\tan ^2\theta
=\frac{h_8^2-h_1^{'2}}{h_1^2-h_8^2},$$ it is clear that $h_1$ and  
$h_1^{'}$
cannot both be less than $h_8.$ Therefore, $h_1^{'}$ should be  
greater than $h_
8,$ and with the PDG value $h_1^{'}=1380\pm 20$ MeV, one is left  
with $h_1^{'}
\approx 1400$ MeV. In this case, since the $h_1^{'}$ lies slightly  
above the
$h_8,$ the pseudovector nonet has a small positive mixing angle (just in 
opposite to the case of the pseudoscalar nonet for which the $\eta  
$ lies
slightly below the $\eta _8=566$ MeV leading to a small negative  
mixing angle).
The above conclusion would change if one (or both) of the  
$h_1,\;h_1^{'}$
appeared to have a mass higher than the value currently adopted by  
PDG.  \\

We close with a short summary of the findings of this work. \\
1. The nonrelativistic constituent quark model shows a simultaneous mass 
degeneracy of the scalar and tensor meson nonets in the $I=1,1/2,$  
and nearly
mass degeneracy in the $I=0,\;s\bar{s}$ channels. \\
2. Simultaneous mass degeneracy of the axial-vector and  
pseudovector nonets in
the $I=1,1/2$ channels is not excluded in this model, but is  
disfavored by
current experimental data. \\
3. The $q\bar{q}$ assignments for the $P$-wave nonets obtained on  
the basis
of the results of the work, are

1 $^1P_1$ $J^{PC}=1^{+-},$ $b_1(1235),$ $h_1(1170),$ $h_1(1400),$  
$K_{1B}$

1 $^3P_0$ $J^{PC}=0^{++},$ $a_0(1320),$ $f_0(980),$ $\;f_0(1500),$  
$\;K_0^\ast
(1430)$

1 $^3P_1$ $J^{PC}=1^{++},$ $a_1(1190),$ $f_1(1285),$ $f_1(1420),$  
$K_{1A}$

1 $^3P_2$ $J^{PC}=2^{++},$ $a_2(1320),$ $f_2(1270),$  
$f_2^{'}(1525),$ $K_2^
\ast(1430)$

\section*{Acknowledgments}
Correspondence of one of the authors (L.B.) with L.P. Horwitz and P. Page
during the preparation of this work is greatly acknowledged.

\end{document}